\documentclass[prb, twocolumn,superscriptaddress,preprintnumbers,a4paper,amsmath,amssymb,showpacs,floatfix]{revtex4}

\usepackage{graphicx}
\usepackage{color}\color[rgb]{0.000,0.000,0.000} %used for font color
\usepackage{amssymb} %maths
\usepackage{amsmath} %maths
\usepackage[normalem]{ulem} %emphasize weiterhin kursiv
\usepackage {ulem} %\emph{Text}: Text wird unterstrichen

\begin{document}

\newcommand{\ba}{BaFe$_{2}$As$_{2}$}
\newcommand{\ca}{CaFe$_{2}$As$_{2}$} 
\newcommand{\sr}{SrFe$_{2}$As$_{2}$}
\newcommand{\bak}{Ba$_{1-x}$K$_{x}$Fe$_{2}$As$_{2}$}
\newcommand{\feas}{Fe$_{2}$As$_{2}$}

\newcommand{\EF}{E$_{F}$}
\newcommand{\ef}{E$_{F}$}
\newcommand{\Tc}{T$_{C}$}
\newcommand{\tc}{T$_{C}$}
\newcommand{\Ts}{T$_{s}$}
\newcommand{\Tn}{T$_{N}$}

\title{Similarities between structural distortions under pressure and chemical doping in superconducting \ba}

\author{Simon A. J. Kimber}
\email[Email of corresponding author:]{simon.kimber@helmholtz-berlin.de}
\affiliation{Helmholtz-Zentrum Berlin f\"ur Materialien und Energie (HZB), Glienicker Stra\ss e 100, D-14109, Berlin, Germany}
\author{Andreas Kreyssig}
\affiliation{Ames Laboratory, US DOE, Iowa State University, Ames, Iowa 5001, USA}
\affiliation{Department of Physics and Astronomy, Iowa State University, Ames, Iowa 50011, USA}
\author{Yu$-$Zhong  Zhang}
\affiliation{Institute f\"ur Theoretische Physik, Goethe-Universit\"at Frankfurt, Max-von-Laue-Stra\ss e 1, 60438 Frankfurt am Main, Germany}
\author{Harald O. Jeschke}
\affiliation{Institute f\"ur Theoretische Physik, Goethe-Universit\"at Frankfurt, Max-von-Laue-Stra\ss e 1, 60438 Frankfurt am Main, Germany}
\author{Roser Valent\'{i}}
\affiliation{Institute f\"ur Theoretische Physik, Goethe-Universit\"at Frankfurt, Max-von-Laue-Stra\ss e 1, 60438 Frankfurt am Main, Germany}
\author{Fabiano Yokaichiya}
\affiliation{Helmholtz-Zentrum Berlin f\"ur Materialien und Energie (HZB), Glienicker Stra\ss e 100, D-14109, Berlin, Germany}
\author{Estelle Colombier}
\affiliation{Ames Laboratory, US DOE, Iowa State University, Ames, Iowa 5001, USA}
\author{Jiaqiang Yan}
\affiliation{Ames Laboratory, US DOE, Iowa State University, Ames, Iowa 5001, USA}
\author{Thomas C. Hansen}
\affiliation{Institute Max von Laue-Paul Langevin, 6 rue Jules Horowitz, BP 156, F-38042, Grenoble Cedex 9, France}
\author{Tapan Chatterji}
\affiliation{JCNS, Forschungszentrum J\"ulich Outstation at Institut Laue-Langevin, BP 156, F-38042, Grenoble Cedex 9, France}
\author{Robert J. McQueeney}
\affiliation{Ames Laboratory, US DOE, Iowa State University, Ames, Iowa 5001, USA}
\affiliation{Department of Physics and Astronomy, Iowa State University, Ames, Iowa 50011, USA}
\author{Paul C. Canfield}
\affiliation{Ames Laboratory, US DOE, Iowa State University, Ames, Iowa 5001, USA}
\affiliation{Department of Physics and Astronomy, Iowa State University, Ames, Iowa 50011, USA}
\author{Alan I. Goldman}
\affiliation{Ames Laboratory, US DOE, Iowa State University, Ames, Iowa 5001, USA}
\affiliation{Department of Physics and Astronomy, Iowa State University, Ames, Iowa 50011, USA}
\author{Dimitri N. Argyriou}
\email[Email of corresponding author:]{argyriou@helmholtz-berlin.de}
\affiliation{Helmholtz-Zentrum Berlin f\"ur Materialien und Energie (HZB), Glienicker Stra\ss e 100, D-14109, Berlin, Germany}

%\marginpar{\scriptsize right}
\date{\today}
%\preprint{}


\maketitle
\textbf{The discovery of a new family of high \Tc\ materials\cite{Kamihara:2008p6110}, the iron arsenides (FeAs), has led to a resurgence of interest in superconductivity. Several important traits of these materials are now apparent, for example, layers of iron tetrahedrally coordinated by arsenic are crucial structural ingredients. It is also now well established that the parent non-superconducting phases are itinerant magnets\cite{QO,stoner,maps,optical}, and that superconductivity can be induced by either chemical substitution\cite{Rotter} or application of pressure\cite{p2}, in sharp contrast to the cuprate family of materials. The structure and properties of chemically substituted samples are known to be intimately linked\cite{PD,nmat}, however, remarkably little is known about this relationship when high pressure is used to induce superconductivity in undoped compounds. Here we show that the key structural features in \ba , namely suppression of the tetragonal to orthorhombic phase transition and reduction in the As-Fe-As bond angle and Fe-Fe distance, show the same behavior under pressure as found in chemically substituted samples. Using experimentally derived structural data, we show that the electronic structure evolves similarly in both cases. These results suggest that modification of the Fermi surface by structural distortions is more important than charge doping for inducing superconductivity in \ba.}

\begin{figure}[tb!]
\begin{center}
\includegraphics[scale=0.35]{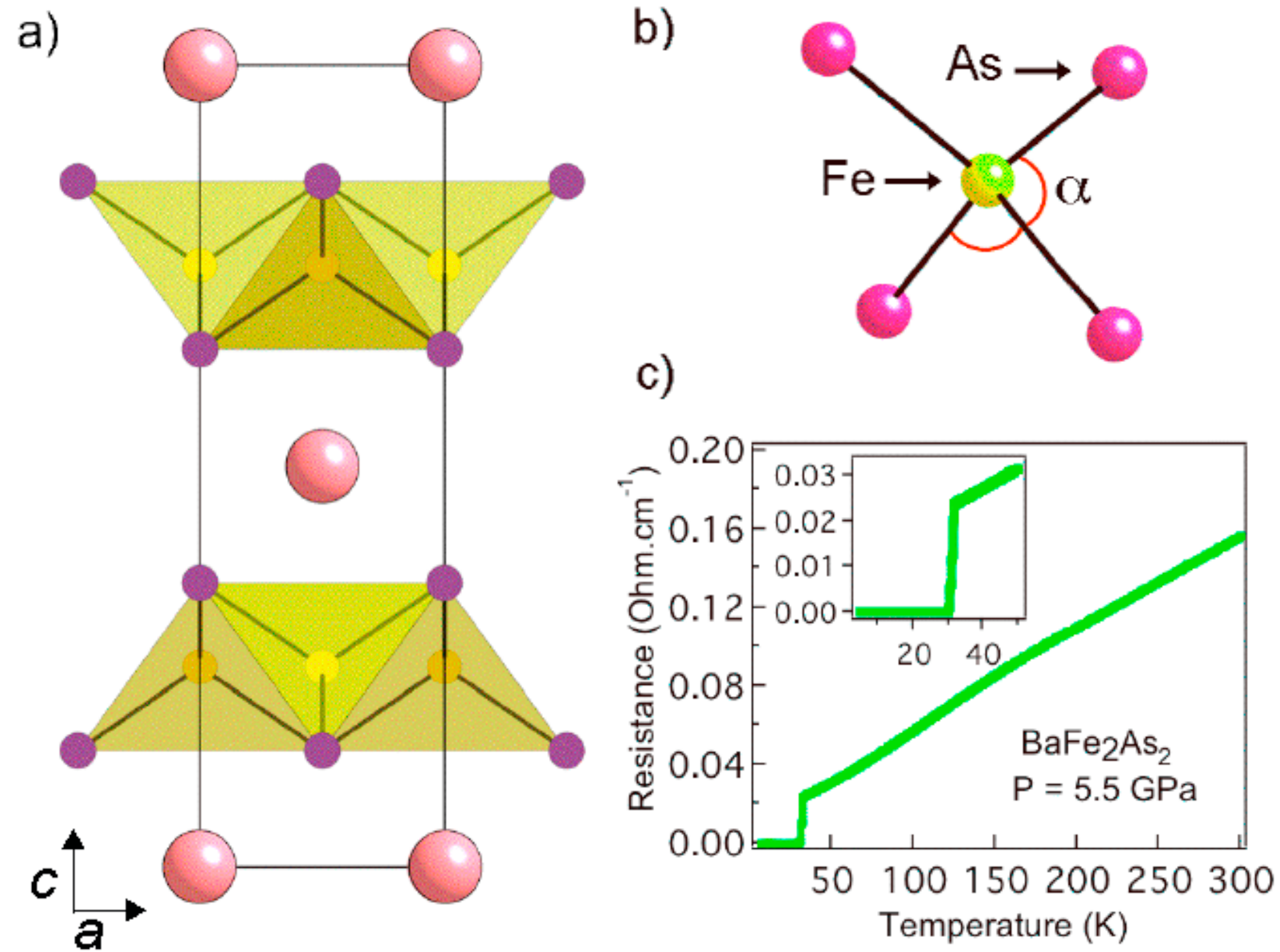}
\caption{\textbf{Crystal structure of \ba, tetrahedral angles and high pressure transport measurements.} A polyhedral representation of the crystal structure of \ba \ is shown in \textbf{a}), iron atoms are shown in yellow, arsenic atoms are purple, barium atoms are pink. The tetrahedral coordination of Fe is shown in \textbf{b}), the angles referred to in the text are highlighted in red. Results of a high pressure resistance measurement at 5.5 GPa on our sample of \ba \ are shown in \textbf{c}), the onset of the transition is at 31 K, zero resistance is achieved at 30.5 K}
\label{}
\end{center}
\end{figure}

In the so-called 'exotic' superconductors, formation of a superconducting condensate of cooper pairs is mediated by magnetic fluctuations. Two classes of materials with transition temperatures at opposite extremes are known. In the $d$-electron cuprate materials, superconductivity emerges from a charge doped antiferromagnetic insulator\cite{review} with T$_{C}$ as high as 160 K. At the other end of the scale, in the heavy fermion superconductors, the groundstate is easily tuned between itinerant spin density wave (SDW) magnetic order and superconductivity\cite{SF} with T$_{C} \sim$1 K. The recently discovered FeAs superconductors bridge this gap with T$_{C}$'s of up to 55 K and a parent phase which has unambigously been shown to be metallic\cite{QO,stoner,maps,optical}. Here we concentrate on the $A$Fe$_{2}$As$_{2}$ (A = Ca, Sr, Ba) family, for which well-characterised single and polycrystalline samples are available (Fig. 1a).  Upon cooling the undoped parent compounds at ambient pressure, a strongly coupled magnetic, structural and electronic transition from tetragonal (T) I4$/mmm$ symmetry to an antiferromagnetically ordered orthorhombic (O) F$mmm$ phase is found\cite{us,them}. Chemical substitution of monovalent cations on the A-site is found to rapidly suppress this transition and superconductivity is found around the phase boundary between the O and T phases\cite{Rotter} with a maximum T$_{C}$ = 38 K. Two effects of chemical substitution can be identified, carrier doping and a steric one, 'chemical pressure' caused by the difference in size between e.g. Ba$^{2+}$ and K$^{+}$. As there is no need to generate charge carriers in the metallic FeAs parent phases, the role of carrier doping in superconductivity is unclear. In contrast, the structural changes which occur as the groundstate is tuned from magnetic to superconducting on chemical substitution are well understood.  In particular T$_{C}$ is found to increase as the As-Fe-As bond angles (Fig.1b) tend to the ideal tetrahedral value of 109.5 deg, and the Fe-Fe nearest neighbour distance is reduced\cite{PD,nmat}. To date however, no investigations have been performed with sufficient detail to resolve the changes in crystal and electronic structure of the superconducting $A$Fe$_{2}$As$_{2}$ compounds under pressure. Consequently, the surprising observation that both doping and pressure induce superconductivity in this family of $d$-electron compounds remains unexplained. 

 \begin{figure}[tb!]
\begin{center}
\includegraphics[scale=0.3]{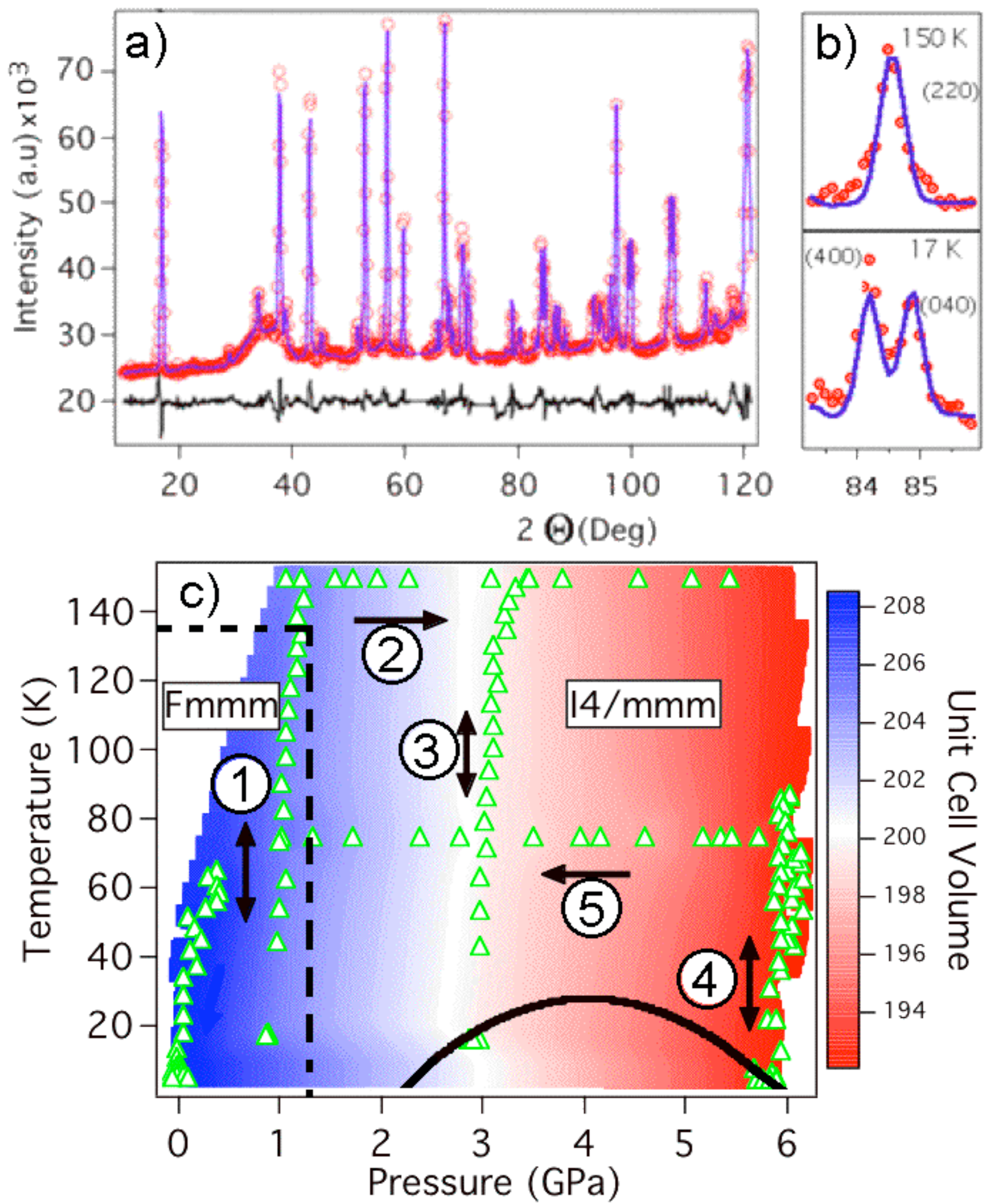}
\caption{\textbf{Results of high pressure neutron diffraction experiment}. The observed, calculated and difference plots for the Rietveld fit to the neutron diffraction profile of \ba \ at 1 GPa and 17 K are shown in \textbf{a}),  the refinement converged with $\chi^{2}$=1.34, R$_{wp}$=0.043. The hump at $\sim$35 deg is from the pressure transmitting medium. Panel \textbf{b}) shows the (220) peak at 150 K in the tetragonal phase (top) and the corresponding (400) and (040) peaks (bottom) in the orthorhombic phase at 17 K. The pressure-temperature phase diagram of \ba \  is shown in \textbf{c}). The cell volume is visualized as a colour map, triangles show experimental data points, arrows show direction of temperature/pressure changes during data collection, dashed line shows approximate T-O phase boundary. Also shown is the schematic superconducting region.} \label{fig3}
\end{center}
\end{figure}

 As shown in Figure 1c, under an applied pressure of 5.5 GPa, our sample of \ba \ shows a sharp drop in resistance at 31 K (midpoint of the transition). Zero resistance is achieved at 30.5 K, a slightly higher temperature than was reported by susceptibility measurements elsewhere\cite{p2}. We note that our observation alone is not absolute confirmation of  'bulk' superconductivity and future pressure dependent measurements such as specific heat may be helpful in this regard. Our high pressure neutron powder diffraction data were collected at a wide range of temperatures and pressures, the high quality of our diffraction data is underlined by the Rietveld fit to the observed neutron diffraction profile of \ba \ at 1 GPa and 17 K shown in Fig. 2a. The (220) peak, which splits at the T-O transition, is also shown at 150 and 17 K in Fig 2b, confirming that our experimental resolution is sufficient to resolve the weak lattice distortion. The pressure/temperature points we measured are shown in Fig. 2c as green triangles, as is the approximate superconducting region.  The isobars and isotherms referred to in the text are marked on the phase diagram with numbers 1 $-$ 5 and were measured in that sequence. After cooling our sample to 150 K we applied a moderate pressure of 1 GPa. On further slow cooling to 17 K (isobar 1), and on warming back to 150 K, we still observed the T-O structural transition at 140 K. At the next highest pressure, 3 GPa, we did not observe this transition on cooling (isobar 3). Also, the orthorhombic phase was recovered at 1.2 GPa upon decreasing pressure at 75 K (isotherm 5).  Both of these facts suggest that the T-O phase line falls sharply. However, a degree of hysteresis at this phase boundary cannot be ruled out, so the phase boundary shown in Fig. 2c should be regarded as schematic in nature. We did not observe coexistence of T and O phases at any pressure/temperature points. The refined unit cell volume across the whole temperature and pressure region studied is shown  in Fig. 2c as an interpolated colour map. The transition to a collapsed phase, as seen\cite{ca1,ca2} in non-superconducting  \ca, would be indicated by a sharp change in the unit cell volume, and is clearly absent in this range of pressures and temperatures (see supplementary information).

  A decrease in the Fe-Fe distance is known to increase T$_{C}$ in the FeAs superconductors\cite{nmat}. We find that this distance ($a$/$\sqrt 2$) in \ba \ decreases linearly (Fig. 3a) up to pressures of 6 GPa. The $a$-lattice parameter from the \bak\ solid solution reported in Ref. 8 is also plotted up to a doping level of x$\sim$0.62, which matches our data well. Although data were plotted such that the maximum T$_{C}$'s reported in refs 7 and 8 agree, we do not imply that pressure is directly proportional to doping. Nevertheless, this comparison does show that the nearest neighbour Fe-Fe distance evolves similarly under chemical or applied pressure as \ba \ is tuned to superconductivity. The change in the $c$-lattice parameter under chemical pressure (an increase of $\sim$7.8 \%) is different to what we find here (4 \% reduction at 6 GPa), which might correlate with the lower superconducting T$_{C}$ and density of states at the Fermi level under pressure (see below), as expected for quasi$-$two dimensional spin fluctuation mediated superconductors\cite{phillip}.

\begin{figure}[tb!]
\begin{center}
\includegraphics[scale=0.5]{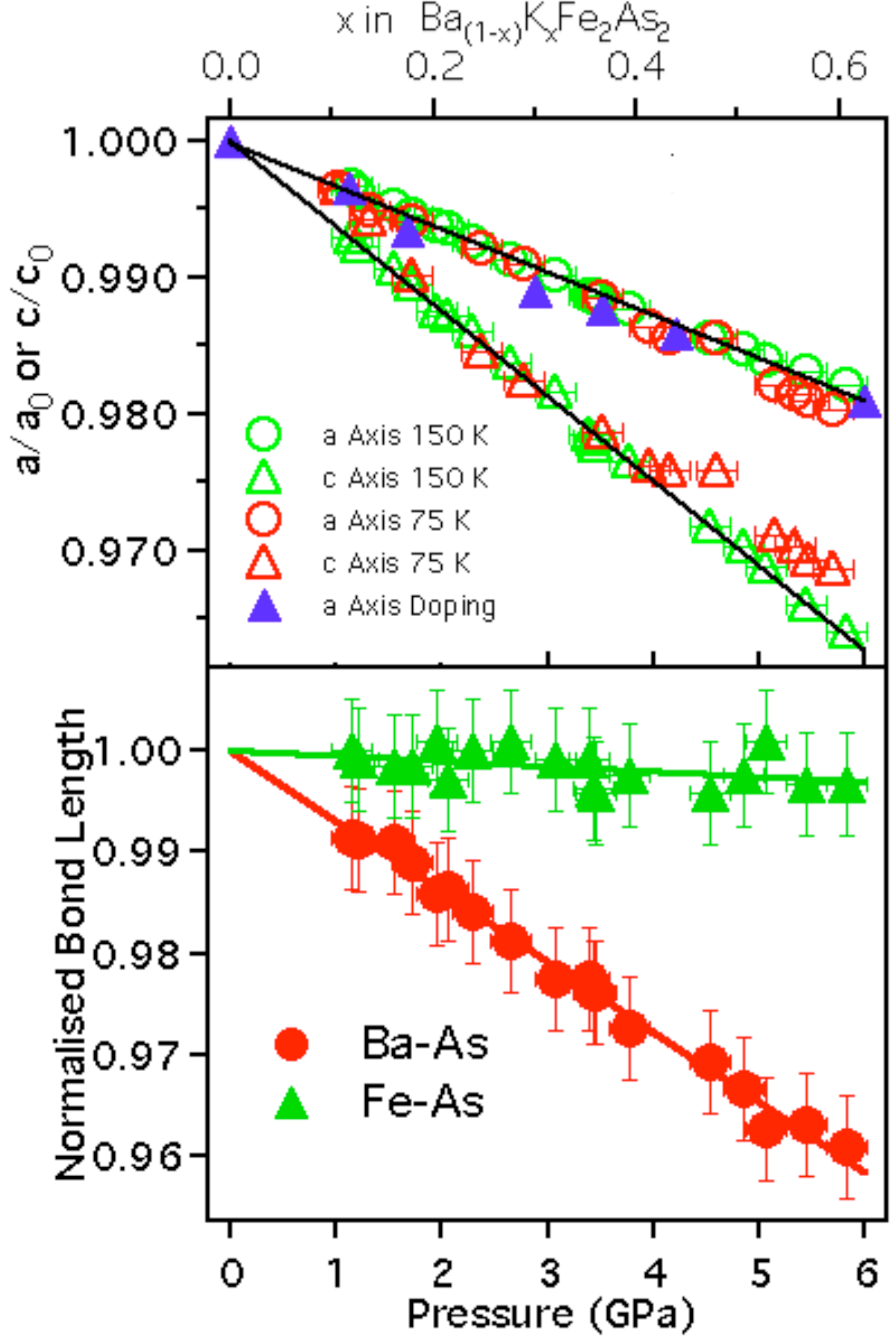}
\caption{\textbf{Structural response of \ba \ to applied pressure}. The pressure dependance of the normalised $a$ and $c$ lattice parameters at 150 K and 75 K, (at 150 K, a$_{0}$ = 3.986(1), c$_{0}$ = 13.058(3) \AA, at 75 K, a$_{0}$ = 3.998(1), c$_{0}$ = 13.0(1) \AA ) are shown in \textbf{a}). We fitted the normalised cell parameters by linear regression, the compressibilities ($k_{a}$=1/$a$(d$a$/d$P$), etc.) are anisotropic, with k$_{a}$ = 3.18(5)x10$^{-3}$ GPa$^{-1}$ and k$_{c}$ = 6.22(5)x10$^{-3}$ GPa$^{-1}$. Also shown is the a-lattice parameter (blue triangles) in the solid solution Ba$_{1-x}$K$_{x}$Fe$_{2}$As$_{2}$ (Ref. 8). The normalised Fe-As and Ba-As bond lengths at 150 K are shown in \textbf{b}) as a function of pressure, the Ba-As compressibility is  k = 6.9(2)x10$^{-3}$ GPa$^{-1}$. Error bars from the pressure calibration and Rietveld refinement are shown in all plots.} \label{fig3}
\end{center}
\end{figure}

\begin{figure}[tb!]
\begin{center}
\includegraphics[scale=0.2]{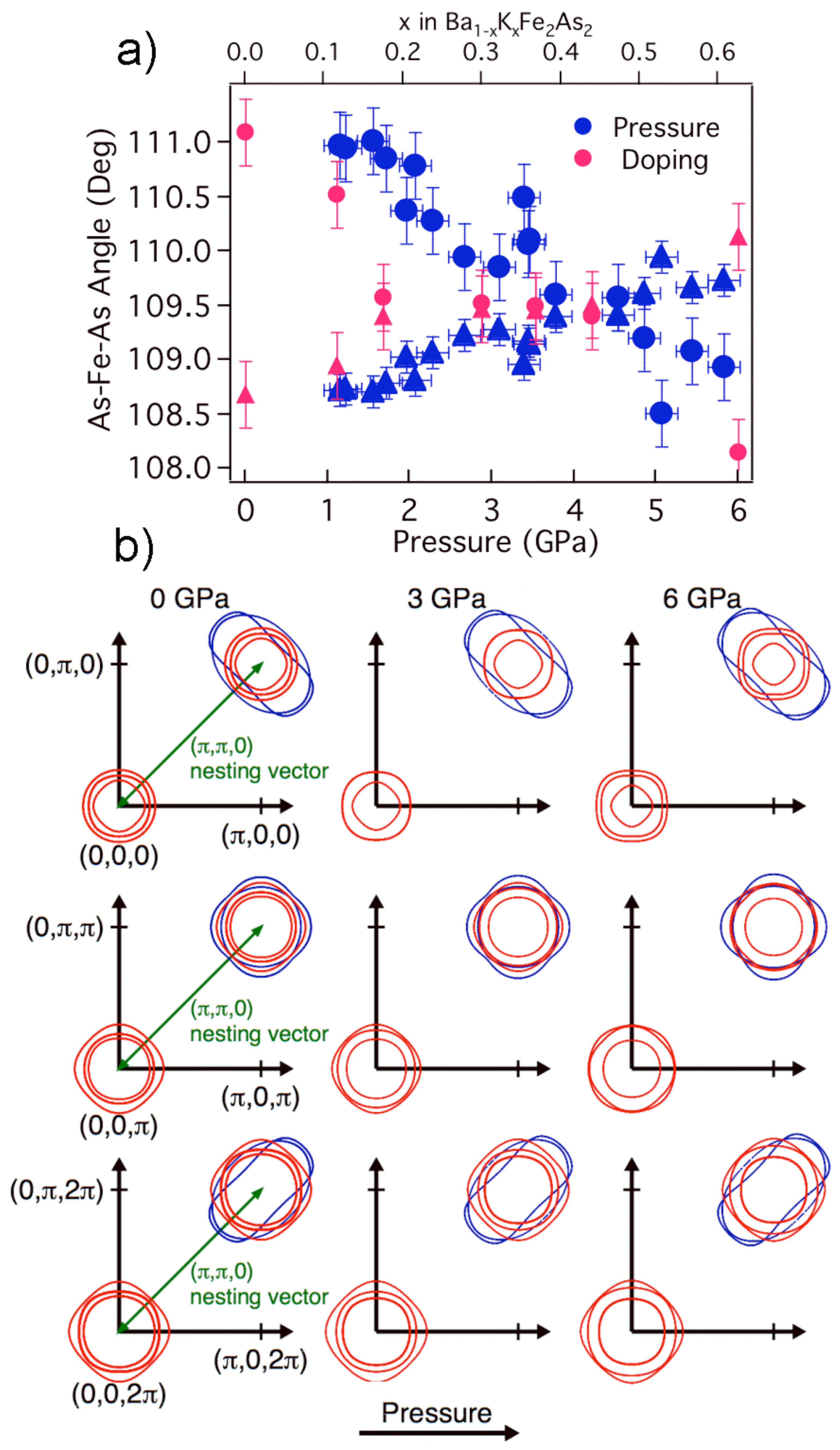}
\caption{\textbf{Pressure dependance of As-Fe-As bond angles and reduction in strength of the Fermi surface nesting at high pressure.} The pressure dependance of the refined As-Fe-As bond angles at 150 K is shown in \textbf{a}), error bars from pressure calibration and Rietveld refinement are shown. The results from the solid solution Ba$_{1-x}$K$_{x}$Fe$_{2}$As$_{2}$ in Ref. 8 are shown as scarlet symbols. \textbf{b}) In each column the intersection of the calculated Fermi surface with planes perpendicular to the c*-direction is shown at three different heights of 0, $\pi$/c, and 2$\pi$/c. Around the origin [0,0], three bands intersect (illustrated by red lines). For a perfect Fermi surface nesting, the translation by the nesting vector [$\pi$/a,$\pi$/b] (in green) should yield a complete overlap with the bands intersecting around the [$\pi$/a,$\pi$/b] point (blue lines).} \label{fig4}
\end{center}
\end{figure}

 The electronic properties of the FeAs superconductors are also sensitively controlled by distortions of the FeAs$_{4}$ tetrahedra\cite{lee}. The degree of distortion is controlled by the As $z$ coordinate, which is the only internal degree of freedom in the tetragonal structure. The quality of our data allowed refinement of this parameter for all pressures and temperatures, and we find that the As $z$ value increases linearly with pressure from 1 GPa to 6 GPa (see Figure S2).  As was seen for the other structural parameters, the refined value on decreasing pressure at 75 K overlies the 150 K values exactly. The extracted Fe-As and Ba-As bond lengths are shown in Fig. 3b as a function of pressure. The Fe-As bond is extremely robust to applied pressure, as was also reported for chemical doping\cite{PD}, whilst the Ba-As bond is found to contract strongly. We also find a striking correlation between the effect of pressure and chemical substitution on the As-Fe-As bond angle, which has been suggested to control the electronic bandwidth in FeAs materials. As shown in Fig. 4a, we show that the As-Fe-As bond angles converge to the ideal tetrahedral value of 109.5 deg as pressure is increased towards the superconducting region, with a possible divergence at higher pressures. The \bak\ solid solution shows the exact same dependance on approaching optimal \Tc, highlighting the importance of this structural parameter in achieving superconductivity. 
 
In summary, our results offer a possible explanation as to why both doping and pressure induce superconductivity in \ba, as the structural changes to the FeAs layer in both cases reduce nesting and hence de-stabilise the SDW groundstate. We have identified a contraction in the nearest-neighbour Fe-Fe distance and a regularisation of the FeAs$_{4}$ tetrahedron as the structural changes responsible and shown that, in contrast with the cuprates, charge doping plays a lesser role. A better analogy for the FeAs superconductors may thus be metallic spin-fluctuation mediated superconductors such as CeCu$_{2}$Si$_{2}$ where pressure and composition tuned competition between itinerant magnetic order and superconductivity are ubiquitous\cite{SF}. However, the changes in individual bond lengths and angles that tune the electronic ground state in these materials are small and difficult to determine due to the low temperatures required (T$_{C}$ $\leq$ 1 K). In addition, the relevant energy scales for the magnetic fluctuations are low and the groundstate is therefore extremely sensitive to variations in sample stoichiometries\cite{stockert}. In contrast, the FeAs superconductors are resistant to disorder, SDW order sets in at high temperature, and the structural changes can be easily resolved by bulk techniques. We anticipate that future high pressure experiments will be key for cleanly exploring the evolution from itinerant magnetism to the exotic superconducting state.

\textbf{Methods}

We synthesised polycrystalline samples of \ba \ by the previously reported method\cite{synth}. Phase purity was checked by powder X-ray diffraction, which showed no impurity phases. The temperature at which \ba \ undergoes antiferromagnetic ordering is extremely sensitive to impurities, we therefore additionally measured the magnetic susceptibity of our sample in a 1 T field using a Quantum Design MPMS. A sharp drop was seen at 143 K, confirming that our sample has close to ideal stoichiometry\cite{Rotter}. An adapted Bridgman cell was used for high pressure transport measurements. The pressure medium was a fluorinert mixture of 1:1 FC70:FC77. The sample (dimensions around 700x150x30 $\mu$m) was measured using the four probe method and the pressure was deduced from the superconducting transition of a piece of lead. Neutron powder diffraction (NPD) profiles were recorded as a function of temperature and pressure using the D20 powder diffractometer\cite{D20} located at the Institut Laue-Langevin with a wavelength of $\lambda$=1.88 \AA. Pressure was applied with a  Paris-Edinburgh pressure cell\cite{cell} equipped with toroidal cubic BN anvils and  a mixture of 4:1 deuterated Methanol:Ethanol was used as pressure transmitting medium. The sample temperature was controlled using a He cryostat similar to that used in previous experiments\cite{cryo}, fast cooling to 70 K was achieved by flooding the cell assembly with liquid N$_{2}$. The cell pressure was determined by adding a small amount of Pb powder to the sample and using the known equation of state of this material\cite{pb}. The NPD data were analysed using the Rietveld method with the program GSAS\cite{GSAS}. We collected data both isothermally  and (approximately) isobarically and typical data collection times were 20 minutes. Various Density functional Theory (DFT) bandstructure calculations have been performed for the doped Ba$_{1-x}$K$_x$Fe$_2$As$_2$ series with results strongly dependent on whether the As positions where relaxed within DFT~\cite{Singh} or were kept fixed as given by experimental observations~\cite{Shein}.  We performed bandstructure calculations for the experimental structures under pressure as well as for the Ba$_{1-x}$K$_x$Fe$_2$As$_2$ solid solution (in the Virtual Crystal Approximation (VCA)) keeping the As position fixed to the experimental values and confirm the remarkable similarities between these two families. We note that previous calculations by Kasanithan $et \ al$\cite{Kasinathan} give a similar result to ours. We employed the full potential linearized augmented plane wave method (FPLAPW) as implemented in the WIEN2k code~\cite{Blaha}. The Perdew-Burke-Ernzerhof (PBE) generalized gradient approximation (GGA) to density functional theory was used.

%\bibliography{prfeaso}

\textbf{Acknowledgments}
We acknowledge the Helmholtz Zentrum Berlin for funding and the Institute Max von Laue-Paul Langevin for access to their instruments. We also thank the high pressure sample environment group of the ILL for technical support. Work at Ames Laboratory was supported by the US Department of Energy$-$ Basic Energy Sciences under Contract No. DE$-$AC02$-$07CH11358. R.V. and H.O.J. thank the DFG for financial support through the TRR/SFB 49 and Emmy Noether programs. S.A.J.K thanks M.V. Kaisheva for a critical reading of the manuscript and D.A. Tennant for helpful discussions.\\

\textbf{Competing financial interests} \\
The authors declare no competing financial interests.

\begin{center}
\textbf{Supporting Information}
\end{center}

\textbf{Collapsed phase in AFe$_{2}As_{2}$ materials with small A-site cations}

In the wider 122 family of compounds, an alternative groundstate is also found for compounds with a small A-site cation such as \ca. This so-called 'collapsed phase' is driven by the formation of As-As bonds between \feas \  layers, and the associated structural deformations\cite{hoffman}.  To assess the extent of such a contribution in \ba, a plot of As-As distances versus the $c/a$ ratio\cite{plot} is shown in S1 for the AFe$_{2}$As$_{2}$ compounds. At 6 GPa the $c/a$ ratio for \ba \ is still larger than \sr \ at ambient pressure. In contrast, the collapsed phase in \ca \ has an As-As distance which approaches that of elemental As (2.52 \AA), inferring a $\sigma$-bonding contribution. \\

\begin{figure}[b!]
\begin{center}
\includegraphics[scale=0.55]{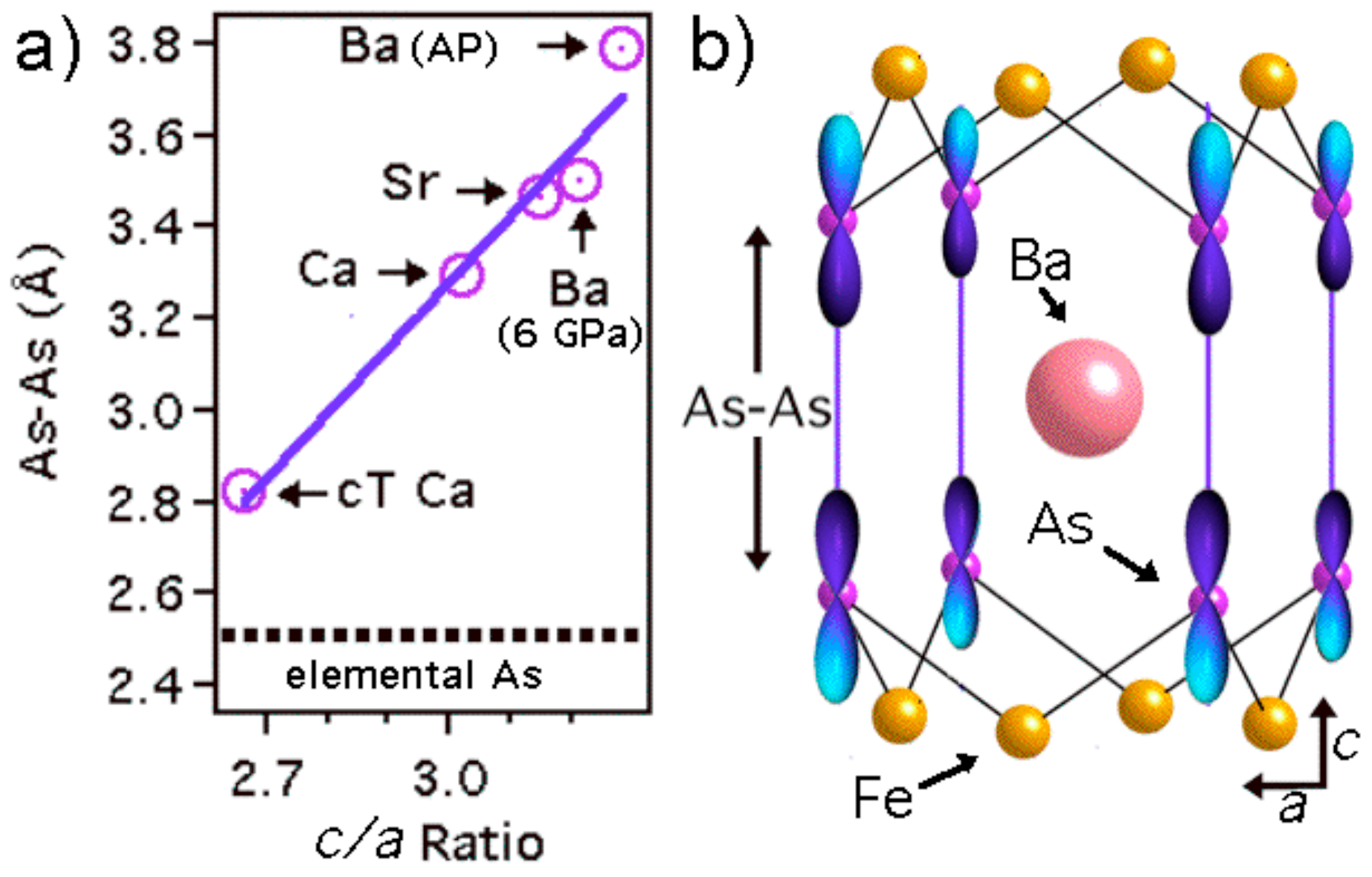}
\caption{\textbf{S1 Details of collapsed phase in AFe$_{2}$As$_{2}$ materials with small A-site cations.}(a) Plot of As-As bond distance against c/a ratio for AFe$_{2}$As$_{2}$ materials, points for high pressure \ca \ and \ba \ are also shown, dashed line shows bond distance in elemental As. (b)As$-$As $\sigma$-bonding pathway in  122 compounds.} \label{fig5}
\end{center}
\end{figure}

\begin{figure}[tb!]
\begin{center}
\includegraphics[scale=0.65]{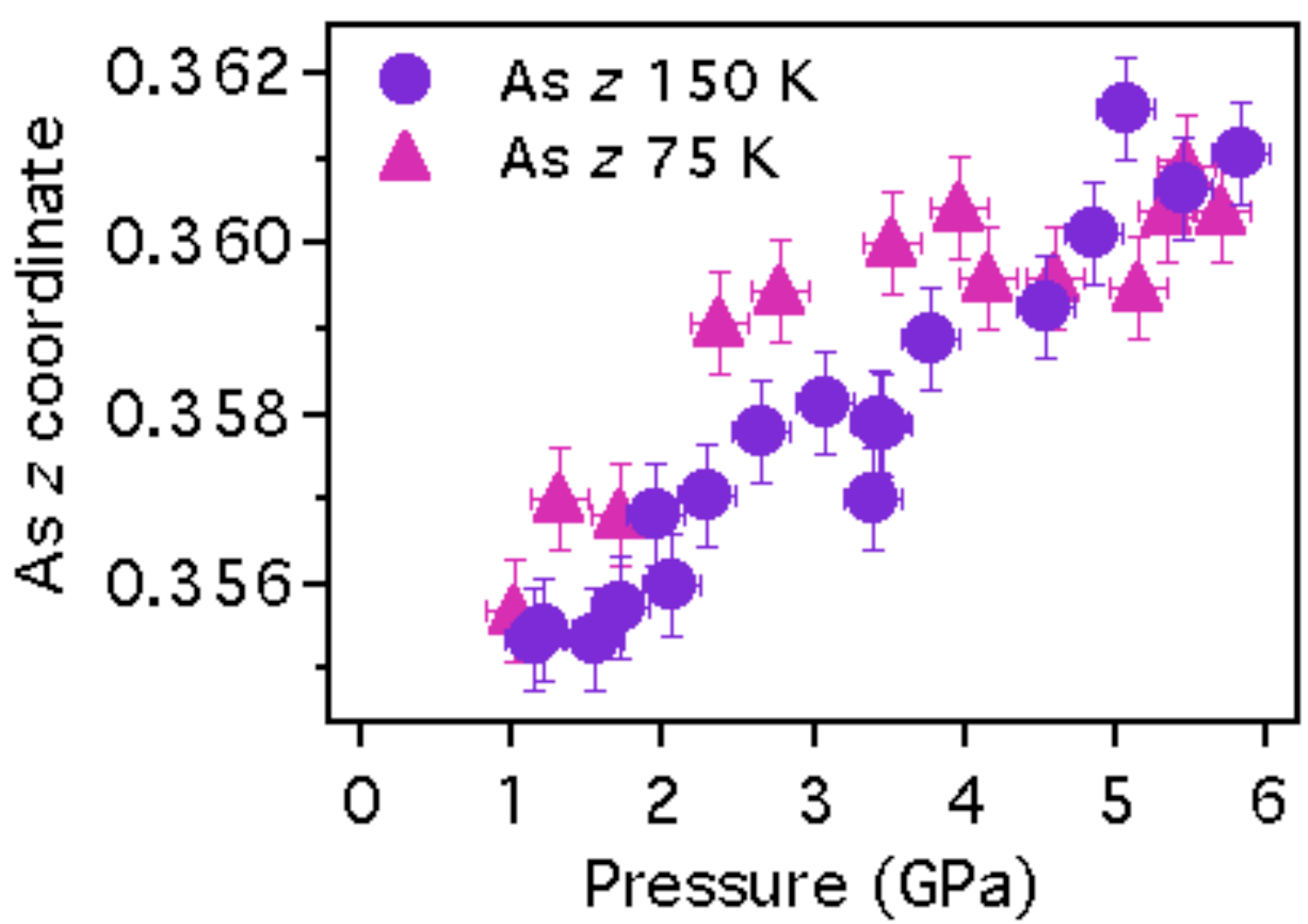}
\caption{\textbf{S2 Refined As $z$-coordinate on increasing pressure at 150 K and decreasing pressure at 75 K from neutron powder diffraction.} Standard uncertainties from Rietveld refinement and error bars from pressure calibration are shown.} \label{fig5}
\end{center}
\end{figure}

\begin{figure}[tb!]
\begin{center}
\includegraphics[scale=0.55]{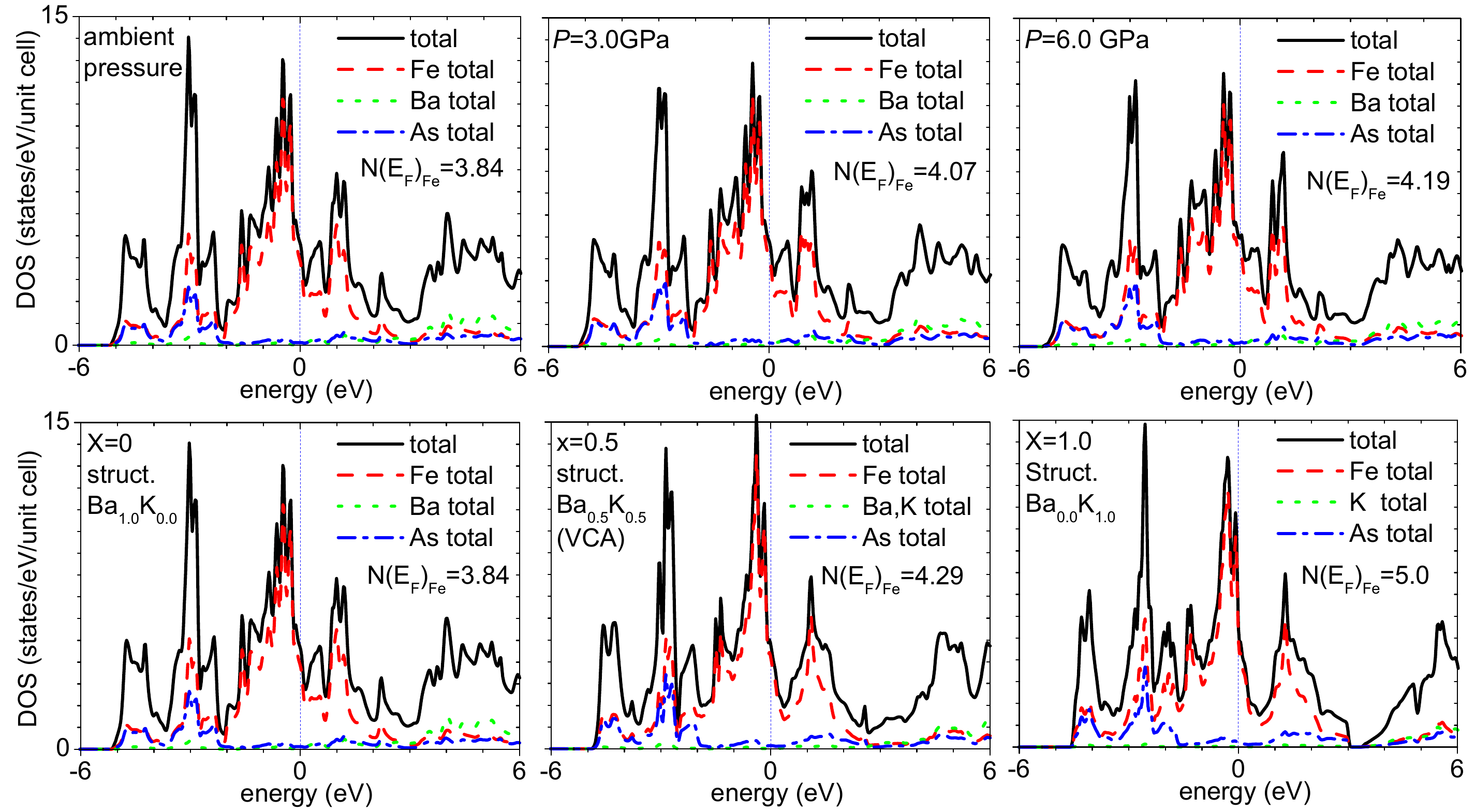}
\caption{\textbf{S4 Calculated Density of States}  Density of states (DOS) for BaFe$_2$As$_2$ at various
  pressures (upper panel) and doping (lower panel). The DOS at E$_{F}$ increases almost linearly as a function of doping and pressure.} \label{fig5}
\end{center}
\end{figure}

\end{document}